\documentclass[aps,prl,preprint,showpacs,superscriptaddress,groupedaddress]{revtex4}  
\usepackage{graphicx}  
\usepackage{dcolumn}   
\usepackage{bm}        
\usepackage{amssymb}   
\usepackage{color}
\newcommand{\dif}{\mathrm{d}}

\newcommand{\p}{\partial}

\DeclareBoldMathCommand{\bV}{V}
\DeclareBoldMathCommand{\bv}{v}
\DeclareBoldMathCommand{\bF}{F}
\DeclareBoldMathCommand{\bg}{g}
\DeclareBoldMathCommand{\bl}{\ell}
\DeclareBoldMathCommand{\bu}{u}
\DeclareBoldMathCommand{\br}{r}
\DeclareBoldMathCommand{\bx}{x}
\DeclareBoldMathCommand{\bz}{z}
\DeclareBoldMathCommand{\bg}{g}
\DeclareBoldMathCommand{\bb}{b}
\DeclareBoldMathCommand{\be}{e}
\DeclareBoldMathCommand{\bs}{s}
\DeclareBoldMathCommand{\bA}{A}
\DeclareBoldMathCommand{\bB}{B}
\DeclareBoldMathCommand{\bC}{C}
\DeclareBoldMathCommand{\bD}{D}
\DeclareBoldMathCommand{\bE}{E}
\DeclareBoldMathCommand{\bI}{I}
\DeclareBoldMathCommand{\bJ}{J}
\DeclareBoldMathCommand{\bM}{M}
\DeclareBoldMathCommand{\bL}{L}
\DeclareBoldMathCommand{\bN}{N}
\DeclareBoldMathCommand{\bP}{P}
\DeclareBoldMathCommand{\bR}{R}
\DeclareBoldMathCommand{\bS}{S}
\DeclareBoldMathCommand{\bU}{U}
\DeclareBoldMathCommand{\bW}{W}
\DeclareBoldMathCommand{\bk}{k}
\DeclareBoldMathCommand{\ba}{a}
\DeclareBoldMathCommand{\bn}{n}
\DeclareBoldMathCommand{\bp}{p}
\DeclareBoldMathCommand{\bq}{q}
\DeclareBoldMathCommand{\br}{r}

\begin{document} 

\title{Flow Shear Suppression of Pedestal Turbulence---A First Principles Theoretical Framework} 

\author{D.R. Hatch} 
\author{R.D. Hazeltine} 
\author{M.K. Kotschenreuther} 
\author{S.M. Mahajan} 
\affiliation{\it Institute for Fusion Studies, University of Texas at Austin, Austin, Texas, 78712} 

\begin{abstract}
	A combined analytic and computational gyrokinetic approach is developed to address the question of the scaling of pedestal turbulent transport with arbitrary levels of $E \times B$ shear.  Due to strong gradients and shaping in the pedestal, the instabilities of interest are not curvature-driven like the core instabilities. By extensive numerical (gyrokinetic) simulations, it is demonstrated that  pedestal modes  respond to shear suppression very much like the predictions of a basic analytic decorrelation theory. The quantitative agreement between the two provides us with a new dependable, first principles (physics based) theoretical framework to predict the efficacy of shear suppression in burning plasmas that lie in a low-shear regime not accessed by present experiments. 
\end{abstract}


 \maketitle 

{\em Introduction.--} 
The interplay between shear flow and turbulence is a central component in the self-organization of wide-ranging fluid and plasma systems.   In neutral fluids, for example, shear flow is a common driver of turbulence (i.e., Kelvin-Helmholtz instabilities).  In contrast, the primary source of turbulence in a typical fusion plasmas is the immense free energy contained in extreme temperature and density gradients. How does the shear flow interact with this class of drift-generated turbulence? 
It is found that, as opposed to its role in hydrodynamic turbulence, the shear flow, in fact, \textit {suppresses} drift turbulence~\cite{terry_99}, and is thought to be the main mechanism underlying the formation and sustenance of the edge transport barrier (called the pedestal) characteristic of the tokamak H-mode~\cite{wagner_82}.  Since the residual turbulence mediates the structure of the pedestal and, consequently, largely determines plasma confinement, it (and its interplay with shear flow) is of central importance to fusion energy.    

The earliest theoretical investigations of shear suppression~\cite{shaing_89, BDT, zhang_92}, which we will call decorrelation theories, predicted reduced fluctuation amplitudes due to the combined advection by background shear flow and self-consistent turbulent flow. Amongst these, the analytic theory of Zhang and Mahajan ~\cite{zhang_92,zhang_93} has compared rather favorably with experimental observations of shear suppression~\cite{boedo_02,schaffner_13}, albeit in experimental setups, perhaps, less challenging than a fusion-relevant H-mode pedestal.

It turns out, however, that the predictions of these basic theories are in striking agreement with gyrokinetic simulations (using the \textsc{Gene} code~\cite{jenko_beta1,goerler_11}) of the pedestal, providing a sound basis for a deep fundamental understanding of the reduction of turbulence by shear flow.  We can, thus, with greater confidence, apply a combination of analytical and numerical approaches to study the scaling of turbulence with flow shear, even in the extreme environment of the H-mode pedestal.  In effect, we will seek a first principles (physics-based) answer to the crucial question: how does turbulence in the pedestal react to a systematic reduction of flow shear rate?  Since all proposed burning plasma machines will lie in the low shear regime, a reliable answer to this question is crucial to the future of fusion energy via Tokamaks operating in H-modes. 
 
After the decorrelation theories of shear suppression (worked out in simple geometry) were proposed, subsequent work emphasized the importance of toroidal effects~\cite{hahm_95,waltz_98,connor_07,barnes_11,staebler_13}.  Toroidal effects are indeed prominent for the conventional instabilities in the \textit{plasma core}. Driven by toroidal curvature (i.e., via resonances with the magnetic drift frequencies), such fluctuations peak in the low magnetic field region of the torus. Interestingly, however, the dynamics of shear suppression manifests differently under conditions characteristic of the pedestal where, due to steep gradients and geometric shaping, the relevant modes are typically not curvature-driven, and consequently are insensitive to toroidal effects ~\cite{kotschenreuther_17}. It is expected, then, that the influence of shear flow on pedestal turbulence  may  be very different from what could be extrapolated from the notions pertinent to the core plasma.  In fact, we reach the surprising conclusion that, despite the substantial complexity and computational challenges involved in pedestal turbulence simulations, the early decorelation theories of shear suppression become highly relevant.  


The importance of this work should be framed as follows.  First, it provides a natural extension of the theory of shear suppression to a pedestal context (perhaps its most important application). Second, building on recent numerical work~\cite{kotschenreuther_17,hatch_17}, it establishes the theoretical underpinnings necessary to understand and therefore to predict/estimate pedestal transport over the transition to lower shear burning plasma regimes. 

{\em Decorrelation Theories---}  
The decorrelation theories of shear suppression  begin with a generic fluid equation of the form
\begin{equation}
\partial_t \xi + \bar{v}(x) \partial_y \xi + \tilde{v}(x,y,t) \partial_x \xi = q(x,y,t),
\label{eqn:fluid_eqn}
\end{equation}
where $x$ is a radial coordinate, $y$ is the corresponding binormal coordinate, $\tilde{v}$ is the fluctuating $E \times B$ velocity, $\bar{v}(x)$ is the macroscopic steady state shear flow, $q$ is a gradient-driven source term, and $\xi$ is a fluid quantity like density or temperature.  Here we analyze the perpendicular temperature fluctuations (i.e. $\xi = \tilde{T}_\perp$---hereafter denoted by $\tilde{T}$) so that Eq.~\ref{eqn:fluid_eqn} may be viewed as a simple analog to Eq. 11 from Ref.~\cite{dorland_93}, which is derived from a moment expansion of the gyrokinetic equations.  The decorrelation theories are based on the so-called \textit{clump theory} described in Ref.~\cite{dupree_72}, and apply basic turbulence closures to solve for properties of the two point correlation function (note that Ref.~\cite{zhang_93} derives similar results via an alternative approach to clump theory).  

For the purpose of this study, we have reproduced a calculation very similar to the original ZM theory~\cite{zhang_92} (due to the close connection, we will refer to our model simply as the ZM theory---details can be found in Appendix A).  The calculation arrives at the following relation describing the reduction of turbulence by shear flow,
\begin{equation}
P(P-\frac{1}{3})(P-1) = \frac{2}{3} W^2 P^{2 \alpha},
\label{eqn:polynomial}
\end{equation}
where $P^{-1}$ represents the reduction in fluctuation amplitudes, 
\begin{equation}
	P^{-1} \equiv \frac{\Delta_{x0}^2}{\Delta_x^2} \frac{\langle \tilde{T}^2 \rangle}{\langle \tilde{T}^2 \rangle_0}, 
\label{eqn:P}
\end{equation}
and $W$ is the normalized shear rate
\begin{equation}
W=\gamma_{E \times B} \tau_{c0}\Theta. 
\label{eqn:W}
\end{equation}
In these expressions, $0$ subscripts denote shear free quantities, $\gamma_{E \times B} = dv/dx$ is the shear rate, the brackets represent ensemble averages (in practice, averages over space and time), $\tau_{c0}$ is the shear-free correlation time, $\Delta_{x,y}$ is the correlation length in the $x,y$ direction (respectively), $\Theta=\Delta_x/\Delta_y$ accounts for anisotropy, and $\alpha$ is a near-unity scaling parameter, which will be described below.  

The ZM theory has two features that distinguish it from other decorrelation theories.  Both are indispensable for the quantitative comparisons that will be described below.  First, the theory is non-asymptotic in shear rate, describing shear suppression seamlessly across the weak and strong shear limits.  Second, and most importantly, it accounts for the fact that fluctuation levels and nonlinear diffusivity are intimately connected and are both sensitively dependent on shear rate.  This is accomplished via an ad hoc expression relating the two: 
\begin{equation}
D = D_* \langle \tilde{T}^2 \rangle^\alpha, 
\label{eqn:DvsT}
\end{equation}
where $D$ is the nonlinear diffusivity, $D_*$ is a constant proportionality factor, and $\alpha$ is the relevant scaling parameter related to the strength of the turbulence ($\alpha \sim 0.5/1.0$ for strong/weak turbulence, respectively).  
In summary, given a shear rate, correlations lengths, and the scaling parameter $\alpha$, Eq.~\ref{eqn:polynomial} predicts the relative suppression of turbulent fluctuation amplitudes.  
In essence, the theory captures the nonlinear decorrelation of turbulence when subject simultaneously to a background shear flow and a self-consistent turbulent flow.  By balancing this decorrelation with a generic gradient drive, an expression is derived for the reduction of turbulence by shear flow.  Notably, the theory neglects parallel dynamics (e.g. Landau damping), zonal flows~\cite{diamond_zf}, toroidal effects, non-local (i.e., global) effects~\cite{lin_02,candy_04,mcmillan_10,goerler_11b}, details of the driving instability, coupling with damped eigenmodes~\cite{terry_06,hatch_11,makwana_14,hatch_16b}, and non-monotonic flow profile variation, all of which are included in our simulations.  Thus, to the extent that simulation and theory agree, it can be concluded that the underlying mechanism of shear suppression in the pedestal is described by a few relatively simple ingredients.  Presently, we make such comparisons.  

\begin{figure}
\includegraphics[scale=0.8]{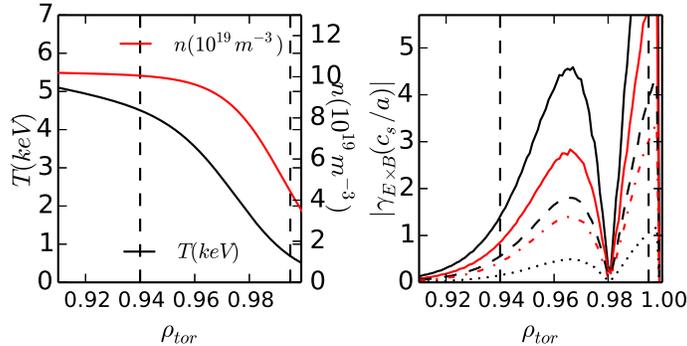}
\caption{\label{fig:profiles} Representative profiles of density and temperature (a) and $E \times B$ shear rate (absolute value) (b).  The dashed vertical lines denote the simulation domain for gyrokinetic simulations.}
\end{figure}

{\em Comparisons between Theory and Simulation---}  
The gyrokinetic \textsc{Gene} simulations described here are designed to include a wide range of relevant pedestal effects while also facilitating clear comparisons with theory.  To this end, we employ an adiabatic electron approximation in order to reduce computational demands and limit the dynamics to the ion temperature gradient (ITG) driven turbulence of interest.  We note that this ITG turbulence is not the dominant pedestal transport mechanism in most present day experiments precisely due to its suppression by shear flow.  The most important fluctuations are likely electron temperature gradient turbulence~\cite{told_08,jenko_09,canik_14,hatch_15,hatch_16}, microtearing modes~\cite{hatch_16,hatch_17}, and low-n (toroidal mode number) magnetic fluctuations~\cite{snyder_09a,yan_11a,diallo_14,diallo_15,laggner_16}---all of which are expected to be much less sensitive to shear flow than ITG.    

Recent related work (Refs.~\cite{kotschenreuther_17,hatch_17}) explores the implications of the expected $\rho_*$ scaling of pedestal flow shear ($\rho_*$ is the ratio of the sound gyroradius to the minor radius $\rho_s / a$). 
The pedestal shear rate is effectively determined by the self-organization of the pedestal by means of force balance between the radial electric field and the pressure gradient.  This force balance, which is well-described by neoclassical theory and well-founded experimentally~\cite{viezzer_13}, results in shear rates that scale linearly with $\rho_*$: $\gamma_{E \times B} \frac{a}{v_{Ti}} \propto \rho_*$ ($v_{Ti}$ is the ion thermal velocity).
Refs.~\cite{kotschenreuther_17,hatch_17} identify two classes of pedestal transport from gyrokinetic simulations---one that scales close to the expected gyroBohm $\rho_*$ scaling , and a second (ITG turbulence) that is small throughout most of the experimentally accessible parameter space but has an unfavorable $\rho_*$ scaling due to its sensitivity to shear flow.  These studies predict the latter mechanism---shear-sensitive ITG turbulence---to be relevant on JET (which has access to the lowest values of $\rho_*$ among active experiments) and to become increasingly important in the transition to low-$\rho_*$ regimes.  To summarize, ITG turbulence in the pedestal is suppressed by the strong shear rates characteristic of present-day experiments.  We are addressing the question of the manner in which it re-emerges as shear rates decrease.  

In order to make comparisons with the ZM theory (Eq.~\ref{eqn:polynomial}), the time- and box-averaged squared temperature fluctuation amplitude, $\langle \tilde{T}^2 \rangle$, and the radial and binormal correlation lengths ($\Delta_x$, and $\Delta_y$) are calculated from simulation data.  The correlation lengths are calculated for temperature fluctuations at the top of the torus where the fluctuation levels peak.  The heat flux, which is embedded in the corresponding temperature moment of the gyrokinetic nonlinearity, provides an appropriate proxy for the nonlinear diffusivity $D$.  The scaling factor $\alpha$ (recall Eq.~\ref{eqn:DvsT}) is extracted from a comparison of $Q_i$ and  $\langle \tilde{T}^2 \rangle$ and lies in the range $\alpha = [0.81,0.97]$ for the cases studied here.  In order to connect the shear rate $W$ (Eq.~\ref{eqn:W}) with its corresponding quantity from the simulations, we use the inverse linear growth rate at $k_y \rho_s=0.1$ as a proxy for the shear-free correlation time $\tau_{c0}$, the standard definition of the $E \times B$ shear rate used in the \textsc{Gene} code~\cite{hatch_15},
and an anisotropy factor $\Theta = \Delta_x/\Delta_y$ defined by the correlation lengths.  One free parameter is used to scale the shear rate and is selected to minimize the discrepancy between simulation and theory.  Encouragingly, this free parameter remains of order unity in all cases studied (varying from $0.58$ to $2.1$).      
The simulations are based on the low $\rho_*$ pedestal setup described in Ref.~\cite{kotschenreuther_17}, which uses JET profile shapes~\cite{leyland_15} in conjunction with ITER geometry and projected ITER pedestal parameters.  Profiles for this base case are shown in Fig.~\ref{fig:profiles} (a).  

\begin{figure}
\includegraphics[scale=0.8]{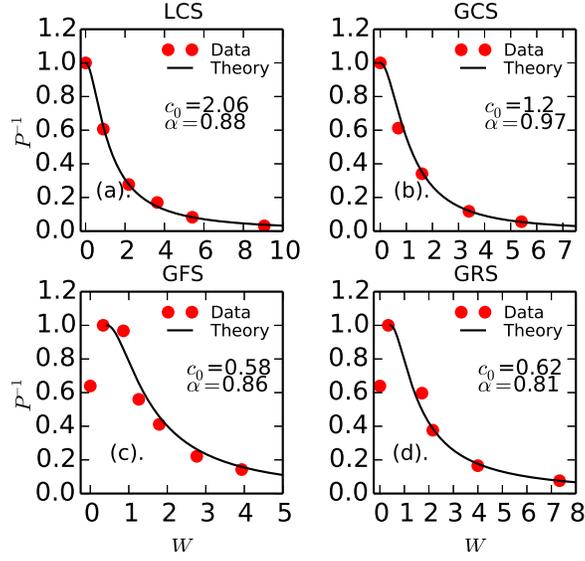}
\caption{\label{fig:shear_suppression_comp} Comparisons between Eq.~\ref{eqn:polynomial} and simulations.  The factor $\alpha$ (Eq.~\ref{eqn:DvsT}) and the free parameter $c_0$ (used to scale the shear rate $W$) are denoted for each case. }
\end{figure}

The extensive simulation campaign described below entails scans of $E \times B$ shear rate for four different scenarios, which are designed to isolate various effects and gauge variation in shear suppression dynamics.  The first case, the \textit{local constant shear} (LCS) case, is designed to match the assumptions of the ZM theory as closely as possible by employing a local approximation (i.e., taking plasma parameters, gradients, and shear rate at a single radial location and neglecting effects from radial profile variation).  A comparison between theory and simulation is shown in Fig.~\ref{fig:shear_suppression_comp} (a); it exhibits a remarkable quantitative prediction of the simulations by the theory.  As demonstrated with the \textit{global constant shear} (GCS) case, which includes self consistent global profile variation (but retains a radially constant shear rate), the theory is robust to the addition of global effects (see Fig.~\ref{fig:shear_suppression_comp} (b)).      The \textit{global full shear} (GFS) case additionally includes non-monotonic flow profiles whose shapes are set by the standard neoclassical expression for the radial electric field (we define the pedestal shear rate to be the radially averaged quantity).  This is particularly significant since it introduces a region of zero shear in the simulation domain (see Fig.~\ref{fig:profiles} (b)), raising the possibility of non-trivial interactions between the turbulence and the flow profile.  We note in this case anomalous behavior in the low shear limit, as shown in Fig.~\ref{fig:shear_suppression_comp} (c), where a discontinuity in $P^{-1}$ between the low shear and the zero shear cases is observed.  Extended simulations targeted at reducing statistical uncertainty produce very minor differences in fluctuation levels, but a persistent ($\sim 20 \%$) decrease in radial correlation length for the low shear case, suggesting that the non-monotonic shear profile acts as a singular perturbation to the length scales.  Consequently, we normalize to the low shear (as opposed to zero shear) case and implement a small corresponding offset in the shear rate.  With this adjustment the simulations for the GFS case also find very good agreement with the theory, as seen in Fig.~\ref{fig:shear_suppression_comp} (c).

{\em Scan of $\rho_*$---}  
While valuable for the purpose of theoretical verification, the three cases examined thus far may be characterized as idealized (and somewhat artificial) setups that exploit the flexibility of our simulation capabilities to independently scan $E \times B$ shear rates.  In an experimental context there is little external control over the shear rate.  As described above, the shear rate is set by the self-organization of the pedestal by means of force balance between the radial electric field and the pressure gradient, resulting in direct proportionality between $\rho_*$ and pdestal shear rates.  Consequently, the most experimentally relevant simulation scenario (called the {\it global rho star} [GRS] case) involves a fully self-consistent scan of $\rho_*$, which holds the pedestal width (in magnetic flux coordinates) fixed along with all other dimensionless parameters (i.e., safety factor q, $\nu_*$, $\beta$).  In this scenario, the $E \times B$ shear profile is determined self-consistently from the density and temperature profiles using the standard neoclassical expression~\cite{hinton_76,landreman_12}.  This scenario involves an additional level of complexity since it conflates the effects of shear suppression with intrinsic $\rho_*$ effects, which are well-known to independently affect turbulence levels (i.e., produce deviations from gyroBohm scaling) as $\rho_*$ is raised above a certain threshold ~\cite{lin_02,candy_04,mcmillan_10,goerler_11b}.  We address this additional complexity with a straightforward modification to the ZM theory.  We assume that finite $\rho_*$ effects are limited to two mechanisms---1) $E \times B$ flow shear, and 2) $\rho_*$ effects manifest in the linear instability drive.  The latter enters the theory while balancing the decorrelation and gradient drive 
\begin{equation}
	\frac{\langle \tilde{T}^2/T_0^2 \rangle}{\tau_c} = \gamma_{lin}(\rho_*)(v_{Ti}/a)\frac{D}{L^2}, 
\label{eqn:gradient_drive}
\end{equation}
where $\gamma_{lin}$ is the $\rho_*$-dependent linear growth rate, $\tau_c$ is the decorrelation time, and $L$ is a macrosopic gradient scale length.  Note that the decorrelation theories (e.g., \cite{BDT,zhang_92}) use Eq.~\ref{eqn:gradient_drive} without the inclusion of the linear growth rate (see Appendix A for additional details).  With this generalization, we find excellent agreement with simulation results, as shown in Fig.~\ref{fig:shear_suppression_comp} (d).

Clearly, the agreement between simulation and theory is substantial in all four cases studied.  This agreement strongly supports a quantitative connection between the simulations and the ZM theory, particularly in light of the following observations: 1) There is good agreement in all four cases studied, which represent substantial variation in physical comprehensiveness.  2) The agreement has been achieved using only a single free parameter, which remains order unity in all cases. 3) As an additional test, we make comparisons while neglecting various aspects of the theory.  The discrepancy increases significantly when the anisotropy facter $\Theta$ and/or the the radial correlation lengths are left out of the theory.          
The results described here suggest that the scaling of pedestal shear suppression of ITG turbulence is determined by the basic ingredients in the decorrelation theory, namely the interplay between turbulent and background advection---both balanced by a gradient drive mechanism.  By extension, the complex physics included in the simulations (zonal flows, Landau damping, damped eigenmodes, global profiles effects, non-monotonic shear profiles, etc.)---which are important for determining absolute fluctuation levels---have little influence on the scaling of turbulence with shear flow. 

\begin{figure}
\includegraphics[scale=0.8]{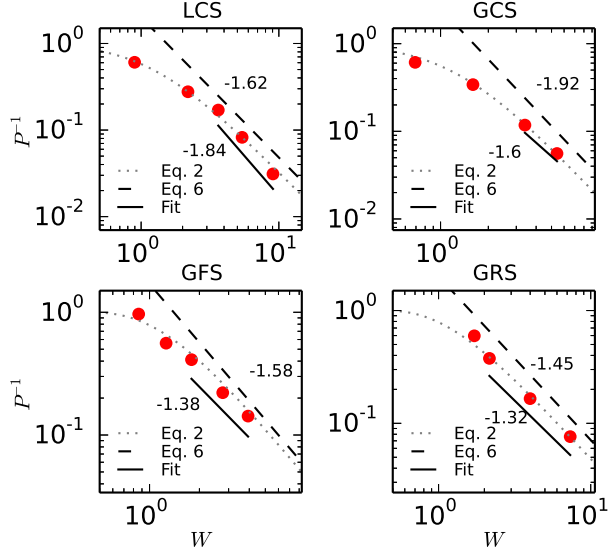}
\caption{\label{fig:strong_shear} Comparisons between Eq.~\ref{eqn:asymptote} and simulations. }
\end{figure}

{\em High Shear Limit---} We now examine the scaling of turbulence in the high shear regime.  The appropriate scaling can be readily derived by taking the $P \gg 1$ limit of Eq.~\ref{eqn:polynomial}
\begin{equation}
	P^{-1} = (2/3)^{1/(2\alpha-3)}W^{2/(2\alpha-3)}.
\label{eqn:asymptote}
\end{equation}
Fig.~\ref{fig:strong_shear} shows a comparison between this high shear scaling and fits to the asymptotic simulation data points.  Note that the asymptotic scaling is strongly dependent on the $\alpha$ factor.  The self-consistent values of $\alpha$ produce a much better match than either the weak turbulence ($\alpha = 1$ translating to $W^{-2}$) or strong turbulence ($\alpha = 0.5$ corresponding to $W^{-1}$) limits.  This high shear scaling has implications for the $\rho_*$ dependence of pedestal transport.  As shown in Fig.~\ref{fig:strong_shear}, turbulence reduction scales roughly as shear rate to the $-3/2$ power in the high shear limit.  Translating the theory-based quantities ($P^{-1}$ and $W$) to more-intuitive quantities (gyroBohm-normalized heat flux $Q/Q_{GB}$ and $\rho_*$) produces (empirically) the rough scaling $Q/Q_{GB} \propto \rho_*^{-2}$, which indicates a transport mechanism that is independent of $\rho_*$.  This scaling is roughly observed for the $\rho_*$ scan described above as well as for similar scans described in Refs.~\cite{kotschenreuther_17,hatch_17}, suggesting its robustness.  This fundamental theoretical prediction must be considered when extrapolating to low $\rho_*$ regimes.  

{\em Discussion---}  
The present study demonstrates that the underlying mechanism of pedestal shear suppression involves a relatively small set of transparent physical ingredients, which are insensitive to parameter variations and difficult to modify. We have, thus, the makings of a first-principles, physics-based theory of turbulence suppression (via shear flow) that can be exploited to estimate/predict turbulent transport in regimes that have not been experimentally probed yet. In fact, ITER and all the future burning plasma experiments fall in the low shear regime not accessed in most current experiments.

We end the paper by pointing out that the simulations/theory suggest several possible routes for controlling turbulent transport in pedestals.  Since the relevant suppression parameter $W$ is the shear rate normalized to the linear growth rate, the most promising route to optimized pedestal performance in low $\rho_*$ regimes is through the minimization of ITG growth rates.  As demonstrated in Refs.~\cite{kotschenreuther_17,hatch_17}, this can be accomplished by at least two mechanisms---1) ion dilution via impurity seeding, and 2) reduction of $\eta_i$ (the ratio of the density and temperature gradient scale lengths) through the separatrix boundary condition.  The latter route will rely heavily on optimization of divertor performance.  

\section{Appendix A---Theory Summary} \label{sec:A}

Here we succinctly outline a derivation of the Zhang-Mahajan (ZM) shear suppression theory~\cite{zhang_92,zhang_93} used in this study.  This derivation is intended to be more accessible than the rigorous but very involved calculation described in~\cite{zhang_93}.  Our simplified derivation first reproduces the orbit equations from Biglari-Diamond-Terry~\cite{BDT} using the standard clump theory~\cite{dupree_72}.  Thereafter, the distinctive elements of the ZM theory are applied, namely a non-asymptotic (in shear rate) treatment of the problem and the use of an \textit{ansatz} relating the nonlinear diffusivity and fluctuation amplitude (described below).  The original ZM theory builds on an alternative set of orbit equations, which stem from an independent approach that does not rely on the standard clump theory.  The model described below exhibits only minor qualitative differences from the original ZM theory.  

 We use coordinates $(x,y)$, where $x$ represents the radial direction and $y$ varies in some direction (other than that of $\bB$) on the magnetic surface.  Parallel gradients, along with such geometrical details as magnetic curvature, are neglected.

A perturbed fluid quantity, such as temperature or density, is denoted by $\xi(x,y,t)$ and assumed to satisfy the equation
\begin{equation}
\label{de1}
\p_{t}\xi + \bar{v}(x)\p_{y}\xi + \tilde{v}(x,y,t)\p_{x}\xi = q(x,y,t)
\end{equation}
Here $\bar{v}(x)$ is the shear flow, a slowly varying equilibrium flow in the $y$-direction, while $\tilde{v}(x,y,t)\ll \bar{v}$ is a turbulent flow. We assume that the turbulent flow is incompressible, and that the variation of $\xi$ on flux surfaces is smaller than its radial variation, so the $x$-component of the turbulent velocity dominates. Rather than trying to solve (\ref{de1}), we derive from it an approximate equation for the correlation function
\[
C_{12} \equiv \langle \xi(x_{1},y_{1},t)\xi(x_{2},y_{2},t)\rangle \equiv \langle \xi_{1}\xi_{2} \rangle
\]
where the angle brackets denote a statistical average. Standard renormalization methods, based primarily on the random phase approximation, yield the evolution equation
\begin{equation}
\label{g1}
(\p_{t} +  \omega_{s} x_{-}\p_{y_{-}} - \p_{x_{-}}(k_{0i}^{2}x_{i-}^{2})D\p_{x_{-}})C_{12}= Q
\end{equation}
where $x_{-} = x_{1} - x_{2}$ is the relative coordinate, $Q$ is the source, $\omega_{s} \equiv  \bar{v}_{c}'$ is the shearing rate, $D$ is a turbulent diffusion coefficient and 
$\bk_{0}$ is a spectral-averaged wave number, related to the width $\Delta$ of an eddy: $
k_{0x,y}  = 1/\Delta_{x,y}$.
The corresponding Green's function $G(\bx_{-},t;\bx_{-}',t') $ satisfies the homogeneous version of (\ref{g1}) with initial data $G(\bx_{-},t;\bx_{-}',t) = \delta(\bx_{-} - \bx_{-}')$.  We are content to study the moments 
\[
M^{ij}(t) \equiv \int \dif \bx \, G(\bx, t; \bx_{0}, 0) x^{i}x^{j}
\]
where $(x^{1},x^{2}) = (x_{-},y_{-})$ and $\bx_{0}$ is some initial value. Integration by parts yields the dynamical moment equations
\begin{eqnarray}
\p_{t}M^{11} &=& 2D k_{\perp}^{2}\left(3M^{11}  + \sin^{2}\theta M^{22}\right) \label{m11} \\
\p_{t}M^{12} &=& \omega_{s}M^{11} + 2Dk_{\perp}^{2}M^{12}   \label{m12} \\
\p_{t}M^{22} &=& 2\omega_{s}M^{12}  \label{m22}
\end{eqnarray}
Here $k_{\perp}^{2} \equiv k_{0x}^{2}$ and $\sin \theta \equiv k_{0y}/k_{0x}$.
Denoting the characteristic time for change in $M_{ij}$ by $\tau_{c}$ and introducing the nominal diffusion time $\tau_{D} \equiv (k_{\perp}^{2}D)^{-1} = \frac{\Delta^{2}}{D}$ we obtain the characteristic equation \begin{equation}
\label{ce}
z(z-2)(z-6) = 4(\omega_{s}\tau_{D})^{2}\sin^{2}\theta
\end{equation}
where $z \equiv \tau_{D}/\tau_{c}$. 
Since the gradients relax through turbulent diffusion, the source for turbulence is measured by $D/L^{2}$.  This observation leads to the estimate
\begin{equation}
\label{pp}
\frac{1}{\tau_c} \left \langle \frac{\tilde{T}^{2}}{T_0^2} \right \rangle = D/L^{2}
\end{equation}
We concentrate on the fastest relaxation rate, $z_{0} = 6$, where the subscript refers to zero shear.  The effects of shear are displayed through the ratio 
\begin{equation}\label{def2}
	\frac{z}{z_{0}} = \frac{\Delta^{2}}{\Delta_{0}^{2}}\frac{\langle \tilde{T}^{2}\rangle_{0}}{\langle \tilde{T}^{2}\rangle} \equiv P
\end{equation}
Following \cite{zhang_92}, we adopt the \emph{ansatz} $D = D_{*}\langle \tilde{T}^{2} \rangle^{\gamma}$, where $D_{*}$ is independent of the turbulence level and scale.  Then (\ref{ce}) becomes
\begin{equation}
\label{pfin}
P\left(P - \frac{1}{3}\right)\left(P -1\right)  = \frac{2}{3}W^{2}P^{2\gamma}
\end{equation}
where $
W = \tau_{c0}\left(\frac{\Delta^{2}}{\Delta_{0}^{2}}\right)^{1-\gamma}
$.
Aside from the numerical details, (\ref{pfin}) agrees with \cite{zhang_93}.

In the case of a self-consistent $\rho_*$ scan, Eq.~\ref{pp} must be modified to account for the intrinsic $\rho_*$ effects manifest in the linear growth rate  
\begin{equation}
\label{}
\frac{1}{\tau_c} \left \langle \frac{\tilde{T}^{2}}{T_0^2} \right \rangle = \gamma_{lin}(\rho_*)(v_{Ti}/a)D/L^{2}.
\end{equation}
This generalization translates in the high shear limit into an alternative normalization for the shear rate: $W$ is now normalized to the $\rho_*$-dependent linear growth rate.


{\em Acknowledgements.--} This research used resources of the National Energy Research Scientific Computing Center, a DOE Office of Science User Facility; the HELIOS supercomputer system at the International Fusion Energy Research Center, Aomori, Japan; and the Texas Advanced Computing Center (TACC) at The University of Texas at Austin.  This work was supported by U.S. DOE Contract No. DE-FG02-04ER54742.

\end{document}